\documentclass[fleqn,usenatbib]{mnras}

\usepackage{newtxtext,newtxmath}

\usepackage[T1]{fontenc}

\DeclareRobustCommand{\VAN}[3]{#2}
\let\VANthebibliography\thebibliography
\def\thebibliography{\DeclareRobustCommand{\VAN}[3]{##3}\VANthebibliography}

\usepackage{graphicx}	
\usepackage{amsmath}	
\usepackage{amssymb}	
\usepackage{bm}

\newcommand{\wv}{\bm{w}}
\newcommand{\Wv}{\bm{W}}
\newcommand{\wvz}{\bm{w}(0)}
\newcommand{\Pv}{\bm{P}}
\newcommand{\pv}{\bm{p}}
\newcommand{\Rv}{\bm{R}}
\newcommand{\rv}{\bm{r}}
\newcommand{\pt}{p_{\mathrm{t}}}
\newcommand{\diff}{\mathrm{d}}

\title[Step-size effect in the TSI method]{Step-size effect in the time-transformed leapfrog integrator on elliptic and hyperbolic orbits}

\author[Long Wang \& Keigo Nitadori]{
Long Wang $^{1,2}$\thanks{E-mail: long.wang@astron.s.u-tokyo.ac.jp}
Keigo Nitadori $^{2}$
\\
 $^{1}$Department of Astronomy, School of Science, The University of Tokyo, 7-3-1 Hongo, Bunkyo-ku, Tokyo, 113-0033, Japan \\
 $^{2}$RIKEN Center for Computational Science, 7-1-26 Minatojima-minami-machi, Chuo-ku, Kobe, Hyogo 650-0047, Japan\\
}

\date{Accepted XXX. Received YYY; in original form ZZZ}

\pubyear{2015}

\begin{document}
\label{firstpage}
\pagerange{\pageref{firstpage}--\pageref{lastpage}}
\maketitle

\begin{abstract}
  A drift-kick-drift (DKD) type leapfrog symplectic integrator applied for a time-transformed separable Hamiltonian (or time-transformed symplectic integrator; TSI) has been known to conserve the Kepler orbit exactly.
  We find that for an elliptic orbit, such feature appears for an arbitrary step size.
  But it is not the case for a hyperbolic orbit: when the half step size is larger than the conjugate momenta of the mean anomaly, a phase transition happens and the new position jumps to the nonphysical counterpart of the hyperbolic trajectory.
  Once it happens, the energy conservation is broken.
  Instead, the kinetic energy minus the potential energy becomes a new conserved quantity.
  We provide a mathematical explanation for such phenomenon.
  Our result provides a deeper understanding of the TSI method, and a useful constraint of the step size when the TSI method is used to solve the hyperbolic encounters.
  This is particular important when an (Bulirsch-Stoer) extrapolation integrator is used together, which requires the convergence of integration errors.
\end{abstract}




\section{Introduction}

The symplectic integrator can conserve Hamiltonian and angular momentum.
Thus it becomes popular in numerical simulations, especially for the long-term evolution of $N$-body systems, such as planetary systems.
However, one bottleneck of the symplectic method is that a constant step size is necessary.
Thus for the systems with strong variations of interaction, such as high-eccentric or hyperbolic Kepler problems, the method becomes inefficient.

One solution is to apply a time transformation to decouple the time step and the integration step \citep[e.g.][]{Hairer1997}.
But it usually results in an inseparable Hamiltonian so that expensive implicit methods has to be applied together.
\cite{Preto1999} and \cite{Mikkola1999} design a special time-transformation function that an explicit method becomes possible. 
Moreover, they found that when the Hamiltonian is described in a logarithmic form, the drift-kick-drift (DKD) type leapfrog integrator can follow the Kepler trajectory exactly with only round-off errors of energy and phase errors of time.
Such a powerful method then becomes popular in the $N$-body codes for simulating collision stellar systems like star clusters, which require an accurate treatment of close encounters and binary orbits \citep[e.g.][]{Mikkola1993,Aarseth2003,Wang2020c,Wang2020d}.
Besides, this method is also known as algorithmic regularization in \cite{Mikkola1999}.
Hereafter we use a shortened name, time-transformed symplectic integrator (TSI), as a reference to the method.

\cite{Preto1999} also explain why the method can exactly follow the Kepler trajectory (see their appendix).
However, we find that the method fails to integrate the hyperbolic orbit when the step size is too large.
In this work, we describe such an issue and mathematically explain the reason and show how to avoid that.

In section~\ref{sec:tsi}, we briefly introduce TSI method using the description style of \cite{Preto1999}.
Then in section~\ref{sec:test} we show numerical tests of a elliptic orbit and a hyperbolic orbit. In the latter case, a phase transition phenomenon can be observed.
Following that, we provide a mathematical explanation in section~\ref{sec:proof}.
In the end, we discuss and summarize our results in section~\ref{sec:discussion}.

\section{The TSI method}
\label{sec:tsi}

In a classical symplectic integrator for a Hamiltonian system, the time step, $\delta t$, is usually used as the integration step.
Thus $\delta t$ must be constant to keep the symplectic property.
In order to construct a symplectic integrator where $\delta t$ can vary, the extended phase-space Hamiltonian can be applied together \citep[e.g.][]{Hairer1997,Preto1999}:
\begin{equation}
  \Gamma(\Wv) = g(\Wv)\left[H(\wv,t) - H(\wvz,0)\right],
  \label{eq:gamma}
\end{equation}
where $H(\wv,t)$ is the standard Hamiltonian, $g(\Wv)$ is time-transformation function and $\Wv$ is the extended phase-space vector.
In this Hamiltonian, time, $t$, is treated as a coordinate and the corresponding momentum, $\pt\equiv -H(\wvz,0)$.
Thus, $\Wv$ contains the standard phase-space vector, $\wv$, and the new pair of ($t$, $\pt$), where $\wv\equiv(\bm{r},\bm{p})$ and $\bm{r}$ and $\bm{p}$ are coordinates and momenta of all particles.

By introducing the new differential variable, $s$, the equation of motion can be described as
\begin{equation}
  \frac{\diff \Wv}{\diff s} = \{ \Wv, \Gamma(\Wv) \},
\end{equation}
where $\{\}$ is Poisson bracket.
Thus, the time step and integration step ($\delta s$) are decoupled.
A constant $\delta s$ can be chosen while $\delta t$ can vary.

A special type of $g(\Wv)$ introduced by \cite{Preto1999},
\begin{equation}
  g(\Wv) = \frac{f(T(\Pv)) - f(-U(\Rv))}{T(\Pv) + U(\Rv)} ,
\end{equation}
leads to a separable $\Gamma$:
\begin{equation}
  \Gamma(\Wv) = f(T(\Pv)) - f(-U(\Rv)),
  \label{eq:gammas}
\end{equation}
where $\Pv\equiv(\pv, \pt)$; $\Rv\equiv (\bm{r},t)$; $\Wv\equiv(\Rv, \Pv)$; the kinetic energy, $T(\Pv) \equiv T(\pv) + \pt$; and the potential energy, $U(\Rv) \equiv U(\rv,t)$.
Thus, an explicit symplectic integrator such as the second-order leapfrog integrator can be used.

\cite{Preto1999} showed that by choosing
\begin{equation}
  f(x) = \log{x},
  \label{eq:logf}
\end{equation}
the Leap-Frog method with a drift-kick-drift (DKD) mode can ensure that the numerical trajectory of a Kepler orbit follows the exact one with only a phase error of time.
Hereafter "TSI" specifically refers to this DKD mode integrator. 

On the other hand, \cite{Mikkola1999} derived a different form of this method (named as a Logarithmic Hamiltonian).
Mathematically, they are equivalent.

\section{Numerical test}
\label{sec:test}

\begin{figure}
  \centering
  \includegraphics[width=0.95\columnwidth]{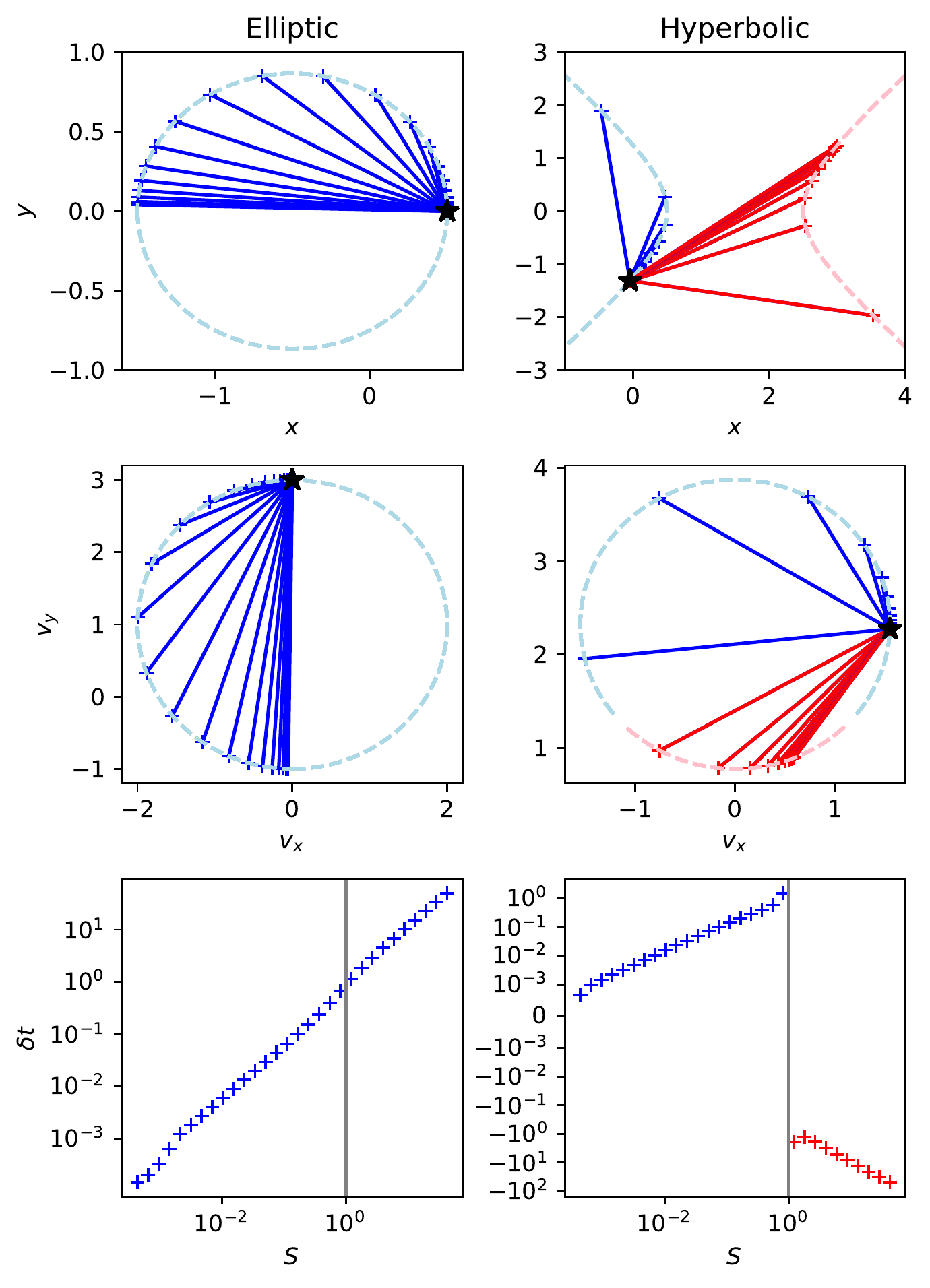}
  \caption{$\bm{r}_1$ and $\bm{v}_1$ after one DKD step by using the TSI method. Each blue or red line represents one choice of integration step, $S$ (Eq.~\ref{eq:S}), which covers a range from $0.001$ to $100$ with an equal interval in the logarithmic scale. The left and right columns show the results of the elliptic and hyperbolic orbits, respectively. The upper panels show $x$-$y$; the middle show $v_x$-$v_y$ and the bottom show $S$-$\delta t$. The dashed light-blue curves show the (physical) elliptic and hyperbolic trajectories. The dashed pink curves show the symmetric counterpart of the hyperbolic trajectory (the upper right panel) and the corresponding velocity curve (the middle right panel). in the hyperbolic case, When $S>1$, $\bm{r}_1$ and $\bm{v}_1$ locate along this nonphysical trajectory and time step $\delta t$ goes backwards.
  }
  \label{fig:xx}
\end{figure}

\begin{table}
    \centering
    \caption{The initial orbital parameters of the elliptic and hyperbolic orbits for the numerical tests. $m_1$ and $m_2$ are masses of components; $a$ is semi-major axis; $e$ is eccentricity; $E$ is eccentric anomaly. The dimensionless units are applied and the gravitational constant is one.}
    \begin{tabular}{cccccc}
    \hline
                   & $m_1$ & $m_2$ & $a$ & $e$ & $E$ \\\hline
        Elliptic   & 1     & 2     & 1   & 0.5 & 0.0 \\
        Hyperbolic & 1     & 2     & -1  & 1.5 & -1 \\\hline
    \end{tabular}
    \label{tab:init}
\end{table}

Since the method can follow the numerical trajectory exactly, it indicates that except for $t$, no truncation error appears on $\bm{r}_1$ and $\bm{v}_1$ for a given $\delta s$, i.e., the error is independent of $\delta s$.
To validate that, in Fig.~\ref{fig:xx}, we show numerical tests of a binary and a hyperbolic encounter by using the TSI method with one DKD step and varying $\delta s$.
The initial orbital parameters are listed in Table~\ref{tab:init}. 
For convenience, the three Euler angles are chosen to be zero so that the orbits locate at the $x$-$y$ plane.
A series of integration step sizes from $0.001$ to $100$ with an equal interval in the logarithmic scale are used for comparison. 
The \textsc{sdar} code \citep{Wang2020c} is used to perform the tests.

We define the scaled integration step as 
\begin{equation}
    S  =  \frac{\delta s}{\mathcal{L}}
  \label{eq:S}
\end{equation}
where $\delta s$ refers to a half step of one DKD step in the TSI method and $\mathcal{L}$ is the conjugate momenta of the mean anomaly:
\begin{equation}
    \mathcal{L}  =  \sqrt{\frac{G (m_1 m_2)^2 |a|}{m_1+m_2}}.
    \label{eq:L}
\end{equation}
Here $a$, $e$, $E$, $m_1$, $m_2$ and $G$ are semi-major axes, eccentricity, eccentric anomaly ($E$), masses of two components and gravitational constant, respectively.
To be convenient, hereafter we use $\bm{r}$ and $\bm{v}$ to represent the relative position and velocity for two-body systems.

Fig.~\ref{fig:xx} suggests that in the elliptic case, the new relative position and velocity after one DKD step, $\bm{r}_1$ and $\bm{v}_1$, are always along the correct trajectory, while the maximum eccentric anomaly change is less than $\pi$.
However, in the hyperbolic case, only when $S<1$, $\bm{r}_1$ and $\bm{v}_1$ can follows the correct trajectory.
When $S>1$, a phase transition appears that $\bm{r}_1$ jumps to the symmetric counterpart of the hyperbolic trajectory and $\bm{v}_1$ jumps to a forbidden region where the value is below the velocity at the infinity position.
On the other hand, the physical time step, $\delta t$, is always positive in the elliptic case.
But in the hyperbolic case, when $S>1$, $\delta t<0$.

\begin{figure}
    \centering
    \includegraphics[width=0.8\columnwidth]{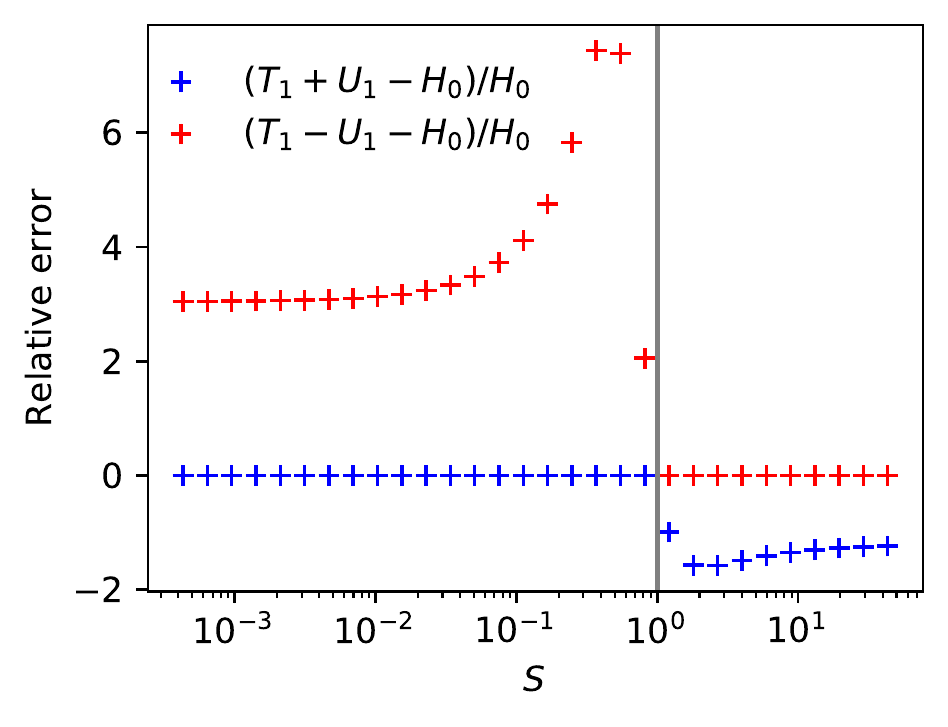}
    \caption{The relative energy error, $(T_1+U_1-H_0)/H_0$, and the pseudo conserved quantity error, $(T_1-U_1-H_0)/H_0$, depending on $S$ for the hyperbolic test.}
    \label{fig:energy}
\end{figure}

Fig.~\ref{fig:energy} show the relative error of the energy and the pseudo conserved quantity. 
When $S<1$, energy is well conserved.
But when $S>1$, energy error is very large and it is not a monotonic function of $S$.
Instead, a new conservation law appears that $T_1-U_1=H_0$.
Thus the integrated result becomes nonphysical.
Nevertheless, this counterpart of the hyperbolic trajectory can represent a physical case where the central force is repulsive.
We discuss the geometrical and physical aspects of this counterpart orbit in \ref{sec:repulsive}.

\begin{figure}
    \centering
    \includegraphics[width=0.8\columnwidth]{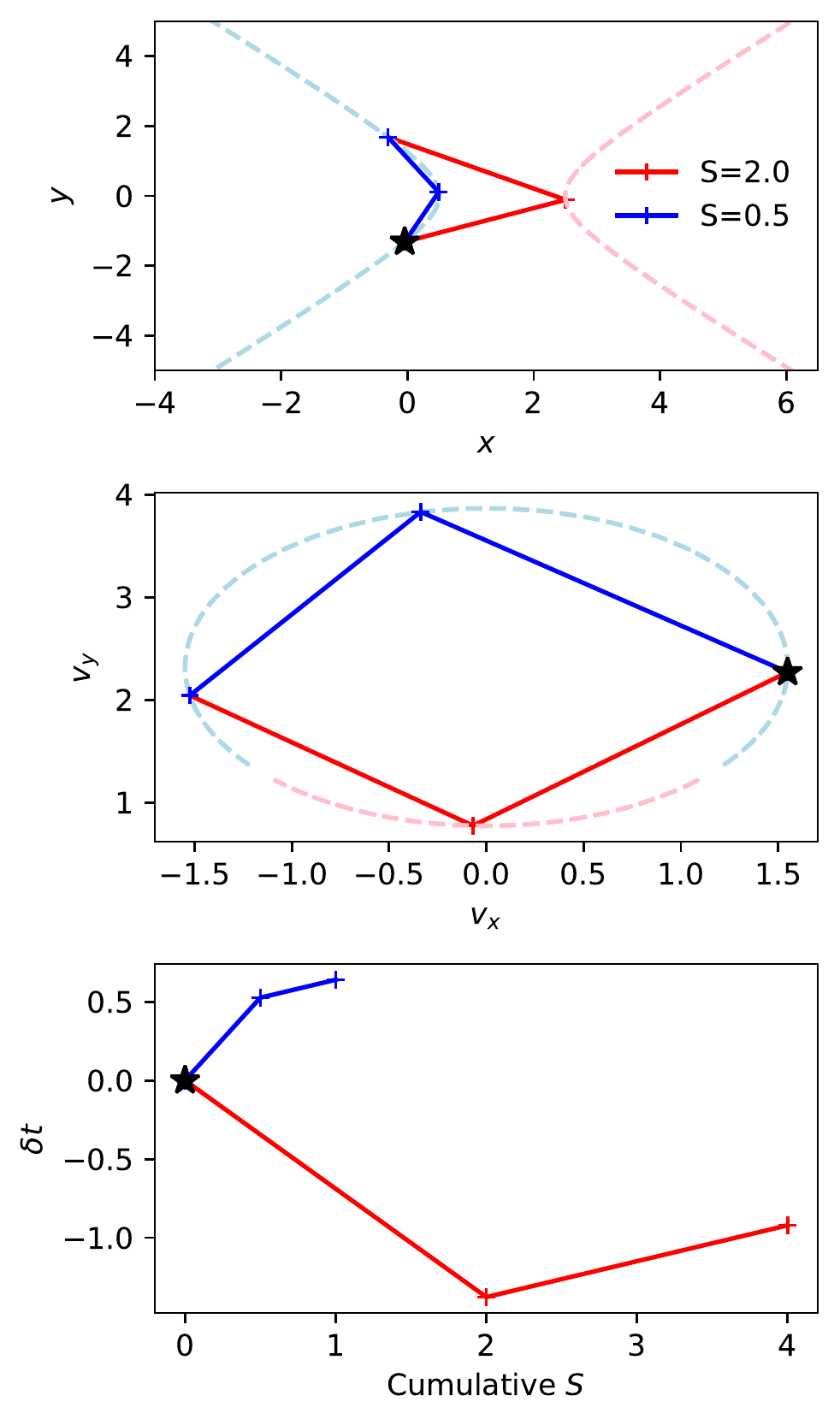}
    \caption{The results of first two DKD steps (DKDKD) with $S=0.5$ and $2.0$ for the hyperbolic test. The plotting style is similar to Fig.~\ref{fig:xx}.}
    \label{fig:dkdkd}
\end{figure}

Such phase transition happens for the first DKD step.
Since $\delta s$ is also large for the following second step, it is worth to investigate that after the second DKD step (DKDKD in total), whether $\bm{r}_2$ and $\bm{v}_2$ can return back to the correct trajectory .
Fig.~\ref{fig:dkdkd} shows a comparison of two step sizes, $S=0.5$ and $2.0$, for the hyperbolic test.
Interestingly, in the case of $S=2$, after two steps, $\bm{r}_2$ and $\bm{v}_2$ not only return back but also overlap the results of $S=0.5$.
If the integration continues, such oscillations of phase transition appears every step.
On the other hand, $\delta t$ cannot return back after the second step.
In the following section, we provide a mathematical explanation for this phase transition.

\section{Explanation}
\label{sec:proof}

\subsection{Elliptic orbit}
We choose a coordinate system that the three Euler angles of the binary is zero.
Thus it is simplified to a two-dimensional problem in the $x$-$y$ plane without losing generality.
Using the orbital elements, the relative position ($\bm{r}$ or $x$ and $y$) and velocity ($\bm{v}$ or $v_x$ and $v_y$), an elliptic orbit can be described by a group of equations:
\begin{equation}
  \left\{
  \begin{aligned}
    x   =& a \left(- e + \cos{\left(E \right)}\right)\\
    y   =& a \sqrt{1 - e^{2}} \sin{\left(E \right)}\\
    v_x =& - \frac{k \sin{\left(E \right)}}{- e \cos{\left(E \right)} + 1} \\
    v_y =& \frac{k \sqrt{1 - e^{2}} \cos{\left(E \right)}}{- e \cos{\left(E \right)} + 1}\\
  \end{aligned}
  \right. ,
  \label{eq:eorbit}
\end{equation}
where $k$ is an orbital speed:
\begin{equation}
    k  =  \sqrt{\frac{G (m_1+m_2)}{|a|}}.\\
  \label{eq:g}
\end{equation}

The first drift of the TSI method can be described as
\begin{equation}
    \bm{r}_{1/2}  =  \bm{r}_{0} + \bm{v}_{0} \frac{\delta s}{T_0-H_0}, \\
    \label{eq:drift}
\end{equation}
where the subscripts $0$ and $1/2$ indicates the initial and the half (first drift) step, and $T_0$ and $H_0$ are initial kinetic and total energy, respectively:
\begin{equation}
  \begin{aligned}
    T_0 & =  \frac{1}{2} \frac{m_1 m_2}{m_1+m_2} \bm{v}_0^2, \\
    H_0 & =  -\frac{G m_1 m_2}{2a}.\\
  \end{aligned}
\end{equation}

Then in the kick step, 
\begin{equation}
    \bm{v}_1  =  \bm{v}_0 + \frac{2 \delta s}{-U_{1/2}} \frac{G (m_1+m_2) \bm{r}_{1/2}}{|\bm{r}_{1/2}|^3}, \\
    \label{eq:kick}
\end{equation}
where $U_{1/2}$ is the potential energy evaluated at the half step:
\begin{equation}
    U_{1/2}  =  -\frac{G m_1 m_2}{|\bm{r}_{1/2}|}. \\
\end{equation}

By using Eq.~\ref{eq:eorbit} as the initial state and doing some mathematical work,
the kicked velocity has the form:
\begin{equation}
  \left\{
  \begin{aligned}
    v_{x,1} & =  \frac{k \left[S^{2} \sin{\left(E \right)} - 2 S \cos{\left(E \right)} - \sin{\left(E \right)}\right]}{S^{2} \left[e \cos{\left(E \right)} + 1\right] + 2 S e \sin{\left(E \right)} - e \cos{\left(E \right)} + 1}\\
    v_{y,1} & =   - \frac{k \sqrt{1 - e^{2}} \left[S^{2} \cos{\left(E \right)} + 2 S \sin{\left(E \right)} - \cos{\left(E \right)}\right]}{S^{2} \left[e \cos{\left(E \right)} + 1\right] + 2 S e \sin{\left(E \right)} - e \cos{\left(E \right)} + 1}  \\
  \end{aligned}
  \right.,
  \label{eq:v1e}
\end{equation}

If the TSI method exactly follows the elliptic trajectory, the kicked velocity should have the form as
\begin{equation}
  \left\{
  \begin{aligned}
    v_{x,1} & =  - \frac{k \sin{\left(E + \delta E \right)}}{- e \cos{\left(E + \delta E \right)} + 1} \\
    v_{y,1} & =  \frac{k \sqrt{1 - e^{2}} \cos{\left(E + \delta E \right)}}{- e \cos{\left(E + \delta E \right)} + 1}\\
  \end{aligned}
  \right.,
  \label{eq:v1se}
\end{equation}
where $\delta E$ is the change of eccentric anomaly.
Eq.~\ref{eq:v1e} and ~\ref{eq:v1se} are equivalent when\footnote{This form is equivalent to the version in \cite{Preto1999} and it provides a better description about the relation between $S$ and $\delta E$. Numerically, it is also more efficient to calculate.}
\begin{equation}
    S = \tan {\left( \frac{\delta E}{2} \right)}.
  \label{eq:Sell}
\end{equation}
Thus, the integration step is directly related to the eccentric anomaly step.
Since the right side of Eq.~\ref{eq:Sell} can have values ranging from $-\infty$ to $\infty$.
Thus $S$ can be an arbitrary real number and the kicked velocity is always on the correct elliptic trajectory with a change of $E$.
When $S=\infty$, $\delta E=\pi$. 
This explains why the numerical result of the elliptic case shown in Fig.~\ref{fig:xx} always passes less than half of the orbit no matter what $S$ it is.

For the second drift, the new position has the form:
\begin{equation}
  \begin{aligned}
    T_{1} & =  \frac{1}{2} \frac{m_1 m_2}{m_1+m_2} \bm{v}_1^2, \\
    \bm{r}_{1} & =  \bm{r}_{1/2} + \frac{\delta s}{T_{1}-H_0}\bm{v}_{1}. \\
  \end{aligned}
  \label{eq:drift2}
\end{equation}
By using Eq.~\ref{eq:v1se}, it can be deduced that
\begin{equation}
  \left\{
  \begin{aligned}
    x_{1} & =  a \left[- S \left( \sin{\left(E \right)} + \sin{\left(E + \delta E \right)} \right) - e + \cos{\left(E \right)}\right] \\
    y_{1} & =  a \sqrt{1 - e^{2}} \left[S \left(\cos{\left(E \right)} + \cos{\left(E + \delta E \right)}\right) + \sin{\left(E \right)}\right] \\
  \end{aligned}
  \right..
\end{equation}
Then using \ref{eq:Sell}, it is not difficult to obtain that
\begin{equation}
  \left\{
  \begin{aligned}
    x_{1} & =  a \left(- e + \cos{\left(E + \delta E \right)}\right) \\
    y_{1} & =  a \sqrt{1 - e^{2}} \sin{\left(E + \delta E \right)} \\
  \end{aligned}
  \right..
\end{equation}
Thus after one DKD step, for any $\delta s$, $\bm{r}$ and $\bm{v}$ are always along the elliptic trajectory.

On the other hand, the physical time changes as\footnote{A sign mistake exists in (A8) of \cite{Preto1999}.}
\begin{equation}
    \begin{aligned}
    \delta t &= \frac{\delta s}{T_0 - H_0} + \frac{\delta s}{T_1 - H_0} \\
             &= \frac{\delta s}{-2H_0} \left[2 - e \cos{\left(E \right)} - e \cos{\left(E + \delta E \right)}\right].
    \end{aligned}
\end{equation}
For any real $\delta s>0$, $\delta t$ is positive and the physical time always advances.

\subsection{Hyperbolic orbit}
\label{sec:hyper}

Similary, the hyperbolic orbit can be described by
\begin{equation}
  \left\{
  \begin{aligned}
    x & =  - a \left(e - \cosh{\left(E \right)}\right) \\
    y & =  - a \sqrt{e^{2}-1} \sinh{\left(E \right)} \\
    v_x & =  - \frac{k \sinh{\left(E \right)}}{e \cosh{\left(E \right)} - 1} \\
    v_y & =  \frac{k \sqrt{e^{2}-1} \cosh{\left(E \right)}}{e \cosh{\left(E \right)} - 1} \\
  \end{aligned}
  \right.,
  \label{eq:horbit}
\end{equation}
where $a<0$ and $e>1$.

After the first kick,
\begin{equation}
  \left\{
  \begin{aligned}
    v_{x,1} & =  - \frac{k \left[S^{2} \sinh{\left(E \right)} + 2 S \cosh{\left(E \right)} + \sinh{\left(E \right)}\right]}{S^{2} \left[e \cosh{\left(E \right)} + 1\right] + 2 S e \sinh{\left(E \right)} + e \cosh{\left(E \right)} - 1} \\
    v_{y,1} & =  \frac{k \sqrt{e^{2} - 1} \left[S^{2} \cosh{\left(E \right)} + 2 S \sinh{\left(E \right)} + \cosh{\left(E \right)}\right]}{S^{2} \left[e \cosh{\left(E \right)} + 1\right] + 2 S e \sinh{\left(E \right)} + e \cosh{\left(E \right)} - 1}\\
  \end{aligned}
  \right..
  \label{eq:v1h}
\end{equation}
When
\begin{equation}
  S = \tanh {\left( \frac{\delta E}{2} \right)},
  \label{eq:Sh}
\end{equation}
$\bm{v}_1$ can follow the correct hyperbolic trajectory with a shift of $\delta E$:
\begin{equation}
  \left\{
  \begin{aligned}
    v_{x,1} & =  - \frac{k \sinh{\left(E + \delta E \right)}}{e \cosh{\left(E + \delta E \right)} - 1} \\
    v_{y,1} & =  \frac{k \sqrt{e^{2} - 1} \cosh{\left(E + \delta E \right)}}{e \cosh{\left(E + \delta E \right)} - 1} \\
  \end{aligned}
  \right..
  \label{eq:v1hs}
\end{equation}

After the second drift, with Eq.~\ref{eq:drift2} and \ref{eq:v1hs},
\begin{equation}
  \left\{
  \begin{aligned}
  x_1 & = - a \left[- S \left(\sinh{\left(E \right)} + \sinh{\left(E + \delta E \right)} \right) + e - \cosh{\left(E \right)}\right] \\
  y_1 & = - a \sqrt{e^{2} - 1} \left[S \left(\cosh{\left(E \right)} + \cosh{\left(E + \delta E \right)}\right) + \sinh{\left(E \right)}\right]
    \end{aligned}
    \right..
\end{equation}
Using Eq.~\ref{eq:Sh}, it can be obtained that
\begin{equation}
  \left\{
  \begin{aligned}
    x_1 & =  - a \left[e - \cosh{\left(E + \delta E \right)}\right] \\
    y_1 & =  - a \sqrt{e^{2} - 1} \sinh{\left(E + \delta E \right)} \\  
  \end{aligned}
  \right..
\label{eq:r1hs}
\end{equation}
Thus, $\bm{r}_1$ is on the correct trajectory.

However, the hyperbolic tangent function has a limited range of $-1$ to $1$.
Thus unlike the elliptic case, $S$ must be less than one, i.e, Eq.~\ref{eq:v1h} and Eq.~\ref{eq:v1hs} are equivalent only when $-1<S<1$.
This explains why a phase transition appears when $S>1$ in the numerical results (Fig.~\ref{fig:xx}).

We can also deduce the orbit for the case of $S>1$.
Eq.~\ref{eq:v1h} shows a symmetric style of coefficients in the polynomial terms of $S$.
This suggests that if a new variable, $\overline{S} = 1/S$, is used, the form of velocity is the same except an opposite sign before "1":
\begin{equation}
  \left\{
  \begin{aligned}
    v_{x,1} & =  - \frac{k \left[\overline{S}^{2} \sinh{\left(E \right)} + 2 \overline{S} \cosh{\left(E \right)} + \sinh{\left(E \right)}\right]}{\overline{S}^{2} \left[e \cosh{\left(E \right)} - 1\right] + 2 \overline{S} e \sinh{\left(E \right)} + e \cosh{\left(E \right)} + 1} \\
    v_{y,1} & =  \frac{k \sqrt{e^{2} - 1} \left[\overline{S}^{2} \cosh{\left(E \right)} + 2 \overline{S} \sinh{\left(E \right)} + \cosh{\left(E \right)}\right]}{\overline{S}^{2} \left[e \cosh{\left(E \right)} - 1\right] + 2 \overline{S} e \sinh{\left(E \right)} + e \cosh{\left(E \right)} + 1}\\
  \end{aligned}
  \right..
  \label{eq:v1hb}
\end{equation}
Now when $S>1$, $\overline{S}<1$, it is possible that 
\begin{equation}
  \overline{S} = \tanh {\left( \frac{\delta \overline{E}}{2} \right)},
  \label{eq:Shb}
\end{equation}
After some mathematical work, $\bm{v_1}$ has the form:
\begin{equation}
  \left\{
  \begin{aligned}
    v_{x,1} & =  - \frac{k \sinh{\left(E + \delta \overline{E} \right)}}{e \cosh{\left(E + \delta \overline{E} \right)} + 1} \\
    v_{y,1} & =  \frac{k \sqrt{e^{2} - 1} \cosh{\left(E + \delta \overline{E} \right)}}{e \cosh{\left(E + \delta \overline{E} \right)} + 1} \\
  \end{aligned}
  \right..
  \label{eq:v1hbs}
\end{equation}
Compared to Eq.~\ref{eq:v1hs}, this form is similar except that the sign before "1" changes.

With this new form of $\bm{v_1}$, after one DKD step, $\bm{r}_{1}$ can be described as
\begin{equation}
  \left\{
  \begin{aligned}
    x_1 & = - a \left[\frac{\sinh{\left(E + \delta \overline{E} \right)} - \sinh{\left(E \right)}}{\overline{S}} + e - \cosh{\left(E \right)}\right] \\
    y_1 & = - a \sqrt{e^{2} - 1} \left[  \frac{\cosh{\left(E \right)} - \cosh{\left(E + \delta \overline{E} \right)}}{\overline{S}}  + \sinh{\left(E \right)} \right] \\
  \end{aligned}
  \right..
\end{equation}
Using Eq.~\ref{eq:Shb}, $\bm{r}_1$ becomes that
\begin{equation}
  \left\{
  \begin{aligned}
    x_1 & =  - a \left[e + \cosh{\left(E + \delta \overline{E} \right)}\right] \\
    y_1 & =    a \sqrt{e^{2} - 1} \sinh{\left(E + \delta \overline{E} \right)} \\
  \end{aligned}
  \right..
\label{eq:x1hbs}
\end{equation}

Comparing Eq.~\ref{eq:r1hs} and \ref{eq:x1hbs}, we can find that $S$ below and above one indicates the two branches of hyperbolic trajectory, where the left branch ($S<1$) is the physical hyperbolic orbit and the right one is a pseudo counterpart as shown in Fig.~\ref{fig:xx}.

At the transition point where $S=1$, after the kick, $\bm{v}_1$ becomes the value at the infinity position.
Thus $T_1-H_0=0$.
In the second drift, from Eq.~\ref{eq:drift2}, division-by-zero occurs and $\bm{r}_1$ becomes infinity.

An interesting point is that in the pseudo branch, the energy conservation law, 
\begin{equation}
     T_1 + U_1 = H_0,   
     \label{eq:tuh0}
\end{equation}
is invalid.
Instead, by using Eq.~\ref{eq:v1hbs} and \ref{eq:x1hbs} a new conservation law can be deduced as
\begin{equation}
    T_1 - U_1 = H_0.
    \label{eq:tuh}
\end{equation}
This explains the phenomenon in Fig.~\ref{fig:energy}.

On the other hand, the physical time changes as 
\begin{equation}
    \begin{aligned}
    &0<S<1:\\
    &\delta t = \frac{s}{2 H_0} \left[e \cosh{\left(E \right)} + e \cosh{\left(E + \delta E \vphantom{\overline{E}} \right)} - 2\right], \\
    &S>1: \\
    &\delta t = \frac{s}{2 H_0} \left[e \cosh{\left(E \right)} - e \cosh{\left(E + \delta \overline{E} \right)} - 2\right]. \\
    \end{aligned}
\end{equation}
When $S>1$, depending on the values of $E$ and $E+\delta \overline{E}$, $\delta t$ can be negative and time goes backwards, as shown in Fig~\ref{fig:xx} and \ref{fig:dkdkd}.

\subsection{Repulsive force}
\label{sec:repulsive}

The classical Hamiltonian for the Kepler orbit has the form of 
\begin{equation}
    H(\wv) = \frac{1}{2} \frac{m_1 m_2}{m_1+m_2} \bm{v}^2  -\frac{G m_1 m_2}{|\bm{r}|} = T(\pv) + U(\rv)
\end{equation}
with the energy conservation law of Eq.~\ref{eq:tuh0}.
When $S>1$, the change of conservation law (Eq.~\ref{eq:tuh}) suggests that the numerical Hamiltonian of the system (which should be the conserved quantity in the symplectic method) becomes different. 

Indeed, when $S>1$, after the kick, $T_1 - H_0 <0$. This indicates that the kinetic part in the Logarithmic-style $\Gamma(\Wv)$ (Eq.~\ref{eq:gammas} and \ref{eq:logf}),
$\log{\left(T_1-H_0\right)}$, has no real solution. 
Thus, when $S>1$, the DKD method is not the solution of this $\Gamma(\Wv)$ at all.
But we can correct the form of $\Gamma(\Wv)$ to match the DKD method, i.e., change the sign of $T(\Pv)$.
Then the new Hamiltonian,
\begin{equation}
    \begin{aligned}
        \overline{\Gamma}(\Wv) & = - \log{\left( - T(\Pv)\right)} - \log{\left(-U(\Rv)\right)} \\
                               & = - \log{\left(H(\wv(0),0) - T(\pv)\right)} - \log{\left(-U(\rv)\right)} \\
    \end{aligned} .
    \label{eq:ngamma}
\end{equation}
The corresponding original (classical) Hamiltonian has the form:
\begin{equation}
    \overline{H}(\wv) = - \left[ \frac{1}{2} \frac{m_1 m_2}{m_1+m_2} \bm{v}^2  + \frac{G m_1 m_2}{|\bm{r}|} \right].
\end{equation}

This Hamiltonian represents a two-body motion with a repulsive central force. 
This also matches the numerical results in Fig.~\ref{fig:xx}.
The counterpart of the hyperbolic trajectory obeys the case where the central force is repulsive \citep{Orti2018}.
Thus, a pseudo conservation law appears.
Such the counterpart only exists in the unbound hyperbolic orbits.
In other words, the two branches of the hyperbolic trajectories can represent the orbits of electron and positron in the Coulomb scattering experiment.

In the counterpart case, a negative sign appears in the front of $\overline{H}(\wv)$.
Thus, $\diff t/\diff s<0$, which results in a negative time step.


\section{Discussion and conclusion}
\label{sec:discussion}

In this work, we show an issue of the TSI method when a hyperbolic orbit is integrated.
For an elliptic orbit, the method can ensure the integrated $\bm{r}$ and $\bm{v}$ are always on the correct trajectory no matter how large the integration step is. 
When step size is positive infinity, the corresponding change of eccentric anomaly is $\pi$, i.e., one DKD step finishes half of the orbit.
But in the hyperbolic case, When the half integration step size is larger than the conjugate momenta of the mean anomaly ($\delta s/\mathcal{L}>1$), the integration fails.
A phase transition appears after one DKD step (Fig.~\ref{fig:xx}) and $\bm{r}$ jumps to the symmetric counterpart of the hyperbolic trajectory while $\bm{v}$ reaches the forbidden region where the value is below the minimum one at the infinite position.
When it happens, the energy conservation law, $T+U=H$, is replaced by a pseudo conservation one, in which $T - U = H$ (see Fig.~\ref{fig:energy}).

To explain the reason, we put the orbital equation into one DKD step and follows the update of $\bm{r}$ and $\bm{v}$. 
A relation between the scaled integration step ($S$) and eccentric anomaly change $(\delta E)$ is found (Eq.~\ref{eq:Sell}, \ref{eq:Sh} and \ref{eq:Shb}).
In the elliptic case, any real number of $S$ can find a corresponding $\delta E$ (Eq.~\ref{eq:Sell}) to ensure the integrated orbit is along the correct trajectory.
But in the hyperbolic case, only when $S<1$, such solution exists (Eq.~\ref{eq:Sh}).
When $S>1$, the orbital equation after one DKD step changes to Eq.~\ref{eq:v1hbs} and \ref{eq:x1hbs}.
This explain the phase transition phenomenon shown in Fig.~\ref{fig:xx}, \ref{fig:energy} and \ref{fig:dkdkd}.
Thus, $S<1$ must be ensured in order to obtain the physical result.
 
In some applications, the TSI method is combined with the Bulirsch-Stoer (BS) method \citep[e.g. the \textsc{archain} code, ][]{Mikkola1999}.
The phase transition behaviour can cause the divergence of energy error (Fig.~\ref{fig:energy}).
Thus the initial step of the BS method should keep $S<1$, otherwise the extrapolation of the integration cannot converge. 
Such problem may appears in an integration of an unstable few-body system.
Usually the total step size of the BS method is large compared to that of the low-order method.
This can be dangerous if the orbit of one inner pair in the system changes to a special hyperbolic one that leads to $S>1$.
Thus the auto adjustment of the step size based on the energy error may not work correctly in the BS method.

In the astronomical applications, the TSI method is used to integrate the orbit of binaries in star clusters \citep[e.g.][]{Aarseth2003,Wang2020d}.
When an supernova appears for one component of a binary, the new formed neutron star or black hole can gain a high natal kick velocity.
Thus, the orbit can suddenly becomes hyperbolic. 
If the step size is not adjusted properly, the phase transition can appear and cause a wrong result in the next integration step. 
This is how this phenomenon is discovered.

One possible application of our analysis is a step-size correction by using Eq. \ref{eq:Sell} or Eq. \ref{eq:Sh} combined with the BS extrapolation method.
Without the correction, each sub-step of the BS method ends on a point of the exact orbit but with different phase errors. 
After the extrapolation, however, this conservation nature is lost.
With the step size corrections, all sub-steps reach an identical point except for the time integral $t(s)$, which is extrapolated to a higher order. 
The energy can conserve exactly.

On the other hand, the new (pseudo) conservation law of Eq.~\ref{eq:tuh} suggests that the DKD method with $S>1$ actually solve a different Hamiltonian (Eq.~\ref{eq:ngamma}), which is a hyperbolic encounter of two bodies with a repulsive force ($G$ is replaced by $-G$). 
Thus, the DKD method can also be used to solve the orbits like the Coulomb scattering of equal-charge particles.


\section*{Acknowledgements}

L.W. thanks the financial support from JSPS International Research Fellow (School of Science, The university of Tokyo).


\section*{Data availability}

The data underlying this article are available in the article and can be generated by using the software \textsc{sdar}, which is available in GitHub, at https://github.com/lwang-astro/SDAR. 



\bibliographystyle{mnras}








\bsp	
\label{lastpage}
\end{document}